\documentclass[11pt]{article}
\usepackage{amsmath}
\usepackage{amsfonts}
\usepackage{amssymb}
\usepackage{graphicx}
\usepackage{bm}
\usepackage{color}
\usepackage{amscd}

\def\bea{\begin{eqnarray}}
\def\eea{\end{eqnarray}}

\begin{document}
\begin{center}
\LARGE {\bf Quasi-local conserved charges in Lorentz-diffeomorphism covariant theory of gravity }
\end{center}

\begin{center}
{H. Adami \footnote{E-mail: hamed.adami@yahoo.com}\hspace{1.5mm} ,
M. R. Setare \footnote{E-mail: rezakord@ipm.ir}\hspace{1mm} \\
{\small {\em  Department of Science, University of Kurdistan, Sanandaj, Iran.}}}\\

\end{center}

\begin{center}
{\bf{Abstract}}\\
 In this paper, using the combined Lorentz-diffeomorphism symmetry, we find a general formula for quasi-local conserved charge of the covariant gravity theories in first order formalism of gravity. We simplify the general formula for Lovelock theory of gravity. Afterwards, we apply the obtained formula on BHT gravity to obtain energy and angular momentum of the rotating OTT black hole solution in the context of this theory.
\end{center}

\section{Introduction}
The concept of conserved charges is a very important matter in gravity theories as well as other physical theories. As we know, the concept of conserved charges of gravity theories is related to the concept of the Noether charges corresponding to the Killing vectors which are admitted by solutions of a theory. There are several approach to obtain mass and angular momentum of black holes solutions  of different gravity theories \cite{2}-\cite{12'}. According to the Arnowitt, Deser and Misner formalism (ADM formalism) \cite{1} one can obtain conserved charges of an asymptotically flat spacetime solution of a general theory of relativity, but this is not a covariant method. The ADM formalism has extended to include asymptotically AdS spacetime solution of Einstein gravity \cite{2}. Deser and Tekin have extended this approach more. By this extension one can calculate the energy of asymptotically dS or AdS solutions in higher curvature gravity models and also in topologically massive gravity model \cite{3}. This method is a covariant formalism which is known as ADT formalism. Thus the ADT formalism is applicable in the general higher curvature and higher derivative theories of gravity \cite{3}-\cite{4}. Another method is the Brown-York formalism \cite{121'} which is based on quasi-local concept, but this approach also is not covariant.
 A general definition of conserved charges in general relativity and other theories of gravity has been proposed in \cite{5}, but this approach is  applicable for asymptotically flat spacetime solutions. This formalism extended to non-covariant theories by Y. Tachikawa \cite{1'}. \\
  Wontae Kim et. al \cite{6} have proposed a way to calculate the conserved charges of all non-asymptotically flat solutions of covariant theories of gravity as well as asymptotically flat one which is based on the concept of the quasi-local conserved charges. This formalism is established on metric formalism of gravity theories and extended to non-covariant theories in \cite{2'}. But we know that one can write down the gravity theories in the first order formalism.
 In the first order formalism of gravity theories, there are some theories which are not only diffeomorphism covariant but also covariant under the local lorentz gauge transformations. The authors in Ref. \cite{7} have combined these two symmetries in an appropriate way and they reached to a new combined symmetry, "Lorentz-diffeomorphism symmetry". In that paper, the authors have shown that the entropy of black holes in the covariant theories of gravity is simply a lorentz-diffeomorphism Noether charge. Recently, this method extended to include non-covariant theories of gravity \cite{8}.\\ Our approach has some similarity with what one can do in the framework of  Poincare gauge theory \cite{a,13,b,c} (see also \cite{d} for recent works in this topic). Poincare gauge symmetry is the symmetry of Poincare gauge theory. In the framework of this theory the translation and Lorentz
symmetry are local. Here we combine the Lorentz gauge transformation with diffeomorphism, then we introduce the total variation under Lorentz-diffeomorphism transformation motivated by  Jacobson, and Mohd \cite{7}. In this case we have a unique transformation, and using it we can obtain conserved charges of a covariant theory of gravity in the first order formalism.\\
 Our aim in this paper is to find quasi-local conserved charges of covariant theories of gravity in the first order formalism. For this purpose, we use of the Lorentz-diffeomorphism symmetry. In this method the quasi-local conserved charges are Lorentz-diffeomorphism Nother charges associated to the Killing vectors which are admitted by considered spacetime. The advantages of this proposal are, first, the quasi-local conserved charges calculated off-shell, second, we can calculate these for the spacetime solutions which are not asymptotically flat (or even AdS).
 \\Our paper is organized as follows. In Sec.2 using Lorentz-Lie (L-L) derivative, we obtain the total variation of the vielbein and the spin-connection. In Sec.3 we consider the most general Lorentz-diffeomorphism invariant Lagrangian $n$-form of a gravity theory and obtain
a general formula for quasi-local conserved charges. In Sec.4 we apply our method on the Lovelock theory in arbitrary dimension and find a general expression for the conserved charges of this theory in any dimension. In Sec.5 using provided formalism we find energy and angular momentum of the rotating OTT black hole solution of BHT gravity. Also, we calculate entropy of this black hole using the general formula for the Chern-Simons-like theories of gravity \cite{8}. Sec.6 is devoted to conclusions and discussions. Sec.6 is devoted to conclusions and discussions.
\section{Lorentz-Lie derivative and total variation}
Consider a $n$-dimensional spacetime. Let $e^{a}_{\hspace{1.5 mm} \mu}$ denote the vielbein, thus we can write the matric as $g_{\mu \nu}=\eta _{ab} e^{a}_{\hspace{1.5 mm} \mu} e^{b}_{\hspace{1.5 mm} \nu}$ where, $\eta _{ab}$ denotes Minkowski metric. Under Lorentz gauge transformation $\Lambda \in SO(n-1,1) $ the vielbein transforms as $\tilde{e}^{a}_{\hspace{1.5 mm} \mu} = \Lambda ^{a}_{\hspace{1.5 mm} b} e^{b}_{\hspace{1.5 mm} \mu}$ so that the spacetime metric under this transformation remains unchanged. In the first order formalism the spin-connection treats as an independent quantity, like the vielbein, and it denotes by $\omega ^{ab}_{\hspace {2 mm} \mu}$. Under Lorentz gauge transformation the spin-connection transforms as $\tilde{\omega}= \Lambda \omega \Lambda ^{-1}+ \Lambda d \Lambda ^{-1}$ so this is not an invariant quantity under considered transformation but we know that under general coordinate transformations the spin-connection transforms as a covariant vector. One can define the vielbein 1-form and the spin-connection 1-form as $ e^{a}= e^{a}_{\hspace{1.5 mm} \mu}dx^{\mu}$ and $ \omega ^{ab}= \omega ^{ab}_{\hspace {2 mm} \mu} dx^{\mu}$, respectively. The Lorentz-Lie (L-L) derivative of the vielbein 1-form is defined as \cite{7}
\begin{equation}\label{1}
  \mathfrak{L}_{\xi} e^{a} = \pounds_{\xi} e^{a} +\lambda ^{a}_{\hspace{1.5 mm} b} e^{b} ,
\end{equation}
where $\pounds_{\xi}$ denotes ordinary Lie derivative along $\xi$ and $\lambda ^{a}_{\hspace{1.5 mm} b}$ generates the Lorentz gauge transformations $SO(n-1,1)$. In general, $\lambda ^{a}_{\hspace{1.5 mm} b}$ is independent from the vielbein and spin-connection and it is a function of space-time coordinates and of the diffeomorphism generator $\xi$. It is straightforward to extend this expression of the L-L derivative for $e^{a}$ to the case for which we have more than one Lorentz index. Now, the total variation under a diffeomorphism, which is generated by a vector field $\xi$, is considered as combination of variation with respect to both of the infinitesimal coordinates and the infinitesimal Lorentz gauge transformation, which is constructed out of diffeomorphism generator $\xi$. Hence, the total variation of the vielbein and the spin-connection are \cite{8}
\begin{equation}\label{2}
  \delta _{\xi} e^{a} = \mathfrak{L}_{\xi} e^{a},
\end{equation}
 \begin{equation}\label{3}
   \delta _{\xi} \omega ^{ab} = \mathfrak{L}_{\xi} \omega ^{ab} -d \lambda ^{ab},
 \end{equation}
respectively. The extra term in \eqref{3}, $-d \lambda ^{ab}$, can be the origin of a non-covariant theory, but here this does not disrupt us because we only focus on the covariant theories.

\section{A general formula for quasi-local conserved charges}
In this section, we try to find a general formula for quasi-local conserved charges of a covariant theory of gravity in the first order formalism. As we know, the curvature 2-form and the torsion 2-form are defined respectively as
\begin{equation}\label{4}
\begin{split}
     & R^{ab}=d \omega ^{ab} + \omega ^{a} _{\hspace{1.5 mm} c} \wedge \omega ^{cb}, \\
     & T^{a} = D e^{a}=d e^{a} + \omega ^{a} _{\hspace{1.5 mm} b} \wedge e^{b} ,
\end{split}
\end{equation}
where, $D$ denotes the exterior covariant derivative. We consider following $n$-form Lorentz-diffeomorphism invariant Lagrangian of a gravity theory
\begin{equation}\label{5}
  L=L(e,T,R,f,h),
\end{equation}
where $ f^{a}$ and $h^{a}$ are Lorentz-diffeomorphism covariant one-form fields. By varying the above Lagrangian with respect to fields, we will have
\begin{equation}\label{6}
  \delta L =\delta e^{a} \wedge E^{(e)}_{a}+\delta \omega ^{ab} \wedge E^{(\omega)}_{ab}+ \delta f^{a} \wedge E^{(f)}_{a}+ \delta h^{a} \wedge E^{(h)}_{a} +d \Theta (\Phi , \delta \Phi).
\end{equation}
Equations of motion are $E^{(e)}_{a}= E^{(\omega)}_{ab}=E^{(f)} _{a} =E^{(h)} _{a} =0$ and $\Theta (\Phi, \delta \Phi)$ is simply the surface term, where $\Phi = \{ e, \omega , f , h\}$.
If this variation is due to a diffeomorphism which is generated by the vector field $\xi$ then, the total variation of Lagrangian \eqref{6} with respect to the diffeomorphism $\xi$ is
\begin{equation}\label{7}
  \delta _{\xi} L =\delta _{\xi} e^{a} \wedge E^{(e)}_{a}+\delta _{\xi} \omega ^{ab} \wedge E^{(\omega)}_{ab}+\delta _{\xi} f^{a} \wedge E^{(f)} _{a} +\delta _{\xi} h^{a} \wedge E^{(h)} _{a} + d \Theta (\Phi , \delta _{\xi} \Phi).
\end{equation}
On the other hand, using \eqref{4}, one can rewrite \eqref{2} and \eqref{3} as follows:
\begin{equation}\label{8}
  \delta _{\xi} e^{a} = D i_{\xi} e^{a} + i_{\xi} T^{a} + (\lambda ^{a b} - i_{\xi} \omega ^{ab}) e_{b} ,
\end{equation}
\begin{equation}\label{9}
  \delta _{\xi} \omega ^{ab} = i_{\xi} R^{ab} + D (i_{\xi} \omega ^{ab} -\lambda ^{a b}) ,
\end{equation}
respectively. Also we have
\begin{equation}\label{10}
\begin{split}
  & \delta _{\xi} f^{a} = D i_{\xi} f ^{a}+ i_{\xi} D f^{a} + (\lambda ^{a b} - i_{\xi} \omega ^{ab}) f_{b} ,\\
  & \delta _{\xi} h^{a} = D i_{\xi} h ^{a}+ i_{\xi} D h^{a} + (\lambda ^{a b} - i_{\xi} \omega ^{ab}) h_{b} ,
\end{split}
\end{equation}
 where $i_{\xi}$ denotes the interior product in $\xi$. By substituting equations \eqref{8}-\eqref{10} into Eq.\eqref{7}, we have
\begin{equation}\label{11}
\begin{split}
   \delta _{\xi} L = & d \left( i_{\xi} e^{a} E^{(e)}_{a}+ i_{\xi} f^{a} E^{(f)}_{a} + i_{\xi} h^{a} E^{(h)}_{a} + (i_{\xi} \omega ^{ab} -\lambda ^{a b}) E^{(\omega)}_{ab} + \Theta (\Phi , \delta _{\xi} \Phi) \right) \\
     & + (\lambda ^{a b} - i_{\xi} \omega ^{ab}) \left( D E^{(\omega)}_{ab} + e_{b} \wedge E^{(e)}_{a} + f_{b} \wedge E^{(f)}_{a} + h_{b} \wedge E^{(h)}_{a} \right) \\
     & - i_{\xi} e^{a} D E^{(e)}_{a} - i_{\xi} f^{a} D E^{(f)}_{a} - i_{\xi} h^{a} D E^{(h)}_{a} + i_{\xi} D f^{a} \wedge E^{(f)}_{a}  \\
     & + i_{\xi} D h^{a} \wedge E^{(h)}_{a} + i_{\xi} T^{a} \wedge E^{(e)}_{a} + i_{\xi} R^{ab} \wedge E^{(\omega)}_{ab}.
\end{split}
\end{equation}
As we expect, the last two lines in the above expression can be rewritten as the following form
\begin{equation}\label{12}
 i_{\xi} e^{a} X _{a} (e,\omega,f) + i_{\xi} f^{a} Y _{a} (e,\omega,f) + i_{\xi} h^{a} Z _{a} (e,\omega,f) .
\end{equation}
So, if we demand that the Lagrangian be invariant under any arbitrary diffeomorphism for arbitrary $e^{a}$, $\omega ^{ab}$ and $f^{a}$, then we must have
\begin{equation}\label{13}
\begin{split}
     & D E^{(\omega)}_{ab} + e_{b} \wedge E^{(e)}_{a}+ f_{b} \wedge E^{(f)}_{a} =0, \\
     & X _{a} (\Phi)=0, \hspace{0.5 cm} Y _{a} (\Phi)=0, \hspace{0.5 cm} Z _{a} (\Phi)=0,
\end{split}
\end{equation}
these equations are Bianchi identities. Considering that the Lagrangian is  a Lorentz-diffeomorphism invariant quantity, then $ \delta _{\xi} L = \mathfrak{L}_{\xi} L = \pounds_{\xi} L = d i_{\xi} L $.
 Therefore, the equation \eqref{11} reduce to the following form
\begin{equation}\label{14}
d J =0,
\end{equation}
where $J$ is an off-shell current density $(n-1)$-form and it expressed as follows:
\begin{equation}\label{15}
J = i_{\xi} e^{a} E^{(e)}_{a} + i_{\xi} f^{a} E^{(f)}_{a} + i_{\xi} h^{a} E^{(h)}_{a} + (i_{\xi} \omega ^{ab} -\lambda ^{a b}) E^{(\omega)}_{ab} + \Theta (\Phi, \delta _{\xi} \Phi) - i_{\xi} L .
\end{equation}
 Since $J$ is a closed form so, by virtue of \cite{9}, we can write $J$ as an exact form, that is $J=dK$. By varying \eqref{15} thus we will find the following expression for the symplectic current
\begin{equation}\label{16}
\begin{split}
   \Omega (\Phi, \delta \Phi , \delta _{\xi} \Phi) = & \delta \Theta (\Phi, \delta _{\xi} \Phi) - \delta _{\xi} \Theta (\Phi, \delta \Phi) \\
     =& d \left( \delta K - i_{\xi} \Theta (\Phi, \delta \Phi) \right) - i_{\xi} e^{a} \delta E^{(e)}_{a} - i_{\xi} f^{a} \delta E^{(f)}_{a}\\
     &  - i_{\xi} h^{a} \delta E^{(h)}_{a} - \delta e^{a} \wedge i_{\xi} E^{(e)}_{a} - \delta f^{a} \wedge i_{\xi} E^{(f)}_{a} - \delta h^{a} \wedge i_{\xi} E^{(h)}_{a}  \\
      & - \delta \omega ^{ab} \wedge i_{\xi} E^{(\omega)}_{ab} - (i_{\xi} \omega ^{ab} -\lambda ^{a b}) \delta E^{(\omega)}_{ab} + \lambda ^{a b} \delta E^{(\omega)}_{ab}.
\end{split}
\end{equation}
 The symplectic current is linear with respect to $\delta _{\xi} \Phi ^{a} $ and $\delta _{\xi} \Phi ^{a} $ become zero when $\xi$ is a Killing vector field. If we demand that $\xi$ be a Killing vector field then $\Omega $ becomes zero. Therefore, in the first order formalism the ADT conserved current is defined as
\begin{equation}\label{17}
\begin{split}
   J_{ADT} = & i_{\xi} e^{a} \delta E^{(e)}_{a} + i_{\xi} f^{a} \delta E^{(f)}_{a} +  i_{\xi} f^{a} \delta E^{(f)}_{a} + (i_{\xi} \omega ^{ab} - \lambda ^{a b}) \delta E^{(\omega)}_{ab} - \lambda ^{a b} \delta E^{(\omega)}_{ab} \\
      & + \delta e^{a} \wedge i_{\xi} E^{(e)}_{a}+ \delta f^{a} \wedge i_{\xi} E^{(f)}_{a} + \delta h^{a} \wedge i_{\xi} E^{(h)}_{a} + \delta \omega ^{ab} \wedge i_{\xi} E^{(\omega)}_{ab}  \\
     = & d \left( \delta K - i_{\xi} \Theta (\Phi, \delta \Phi) \right) .
\end{split}
\end{equation}
We know that the relation between the ADT conserved current and the ADT conserved charge is $ J_{ADT}= d Q_{ADT} $, so the ADT conserved charge is given by
\begin{equation}\label{18}
  Q_{ADT} = \delta K - i_{\xi} \Theta (e,\omega ; \delta e, \delta \omega).
\end{equation}
By taking into account the one-parameter path in the solution space then, the conserved charge associated to the Killing vector field $\xi$ is defined as follows:
\begin{equation}\label{19}
  Q (\xi)= c \int_{0}^{1} ds \int_{\Sigma} Q_{ADT}(e,\omega | s ) ,
\end{equation}
where $c$ is a normalization factor and $\Sigma $ is a time-like $(n-2)$-surface, also, $s=0$ and $s=1$ are correspond to the background solution and the interested solution, respectively. By substituting \eqref{18} into \eqref{19}, we obtain the following expression for the conserved charge associated to the Killing vector field $\xi$:
\begin{equation}\label{20}
  Q (\xi)= c \int_{\Sigma} \left( \Delta K - i_{\xi} \int_{0}^{1} ds \Theta (e,\omega | s ) \right) .
\end{equation}
where $\Delta K =K_{S=1}(\xi)-K_{S=0}(\xi)$.
In this way, we find an expression for the conserved charge which is not depends on  $(n-2)$-surface $\Sigma $. So, we can choose $\Sigma $ everywhere and then we obtain the conserved charges of spacetimes which are not asymptotically flat nor even AdS. \\
As we have mentioned in introduction, our approach has some similarity with what have done in the framework of Poincare gauge theory \cite{a,13,b,c} (see also \cite{14}-\cite{16}, where the authors have studied the conserved charges in the framework of Poincare gauge theory). The difference between this approach and the Poincare gauge theory, is that we combined the Lorentz gauge transformation with diffeomorphism and introduced the total variation under Lorentz-diffeomorphism transformation. In this way, we have a unique transformation and using it we can obtain conserved charges of a covariant theory of gravity in the first order formalism. This method  provides a derivation of the entropy formula for black hole solutions in gravity theories defined by a Chern-Simons gravitational action in $3D$ \cite{8}. It was pointed out in  reference \cite{7} that the derivation of the classical Wald formula for entropy is problematic in the first order formalism using the spin connection. But by introduction the Lorentz-Lie derivative, according to our approach  one can overcome this difficulty.

\section{Quasi-local conserved charges of the Lovelock gravity}
In this section, we apply the above procedure on the Lovelock theory in arbitrary dimension and we will find a general expression for the conserved charges of this theory in any dimension. This theory was first proposed by D. Lovelock \cite{4'} which have the same degrees of freedom as general relativity and it is ghost free. The Lovelock Lagrangian is given by \cite{10}.
\begin{equation}\label{21}
  L(e,R)= \sum_{p=0}^{[n/2]} \alpha _{p} L^{(p)},
\end{equation}
where
\begin{equation}\label{22}
  L^{(p)}= \varepsilon _{a_{1} \cdots a_{n}} R^{a_{1}a_{2}} \wedge \cdots \wedge R^{a_{2p-1}a_{2p}} \wedge e^{a_{2p+1}} \wedge \cdots \wedge e^{a_{n}}.
\end{equation}
In the above Lagrangian, the $\alpha _{p}$ are arbitrary dimensionful coupling constants and $[x]$ denotes the integer part of $x$. Because the procedure which leads to Eq.\eqref{20} is linear in $L$ then we consider $L^{(p)}$. \\
After some calculations, for the Lagrangian Eq.\eqref{21}, we find
\begin{equation}\label{23}
E^{(e)}_{a_{n}} = \frac{\partial L^{(p)}}{\partial e^{a_{n}} } , \hspace{1.5 cm} E^{( \omega )}_{a_{1} a_{2}} = D \frac{\partial L^{(p)}}{\partial R^{a_{1} a_{2} } },
\end{equation}
\begin{equation}\label{24}
  \Theta ^{(p)} (e,\omega ; \delta \omega) = \delta \omega ^{a_{1} a_{2}} \wedge \frac{\partial L^{(p)}}{\partial R^{a_{1} a_{2} } },
\end{equation}
where
\begin{equation}\label{25}
  \begin{split}
     \frac{\partial L^{(p)}}{\partial e^{a_{n}} } = & (n-2p) \varepsilon _{a_{1} \cdots a_{n}} e^{a_{n-1}} \wedge \cdots \wedge e^{a_{2p+1}} \wedge R^{a_{1}a_{2}} \wedge \cdots \wedge R^{a_{2p-1}a_{2p}}, \\
     \frac{\partial L^{(p)}}{\partial R^{a_{1} a_{2} } } =  & p \hspace{1 mm} \varepsilon _{a_{1} a_{2} \cdots a_{n}} R^{a_{3} a_{4}} \wedge \cdots \wedge R^{a_{2p-1}a_{2p}} \wedge e^{a_{2p+1}} \wedge \cdots \wedge e^{a_{n}} .
  \end{split}
\end{equation}
By substituting equations \eqref{22}, \eqref{23} and \eqref{24} into Eq.\eqref{15}, we can read off $K^{(p)}$ as follows:
\begin{equation}\label{26}
  K^{(p)}= (i_{\xi} \omega ^{a_{1} a_{2}} - \lambda ^{a_{1} a_{2}}) \frac{\partial L^{(p)}}{\partial R^{a_{1} a_{2} } }.
\end{equation}
Now, one can calculate the contribution of $p$-term in the conserved charge $Q^{(p)}(\xi)$ using Eq.\eqref{19} and then the conserved charge is $Q(\xi) = \sum_{p=0}^{[n/2]} \alpha _{p} Q^{(p)}(\xi)$. \\
 Thus what we found here is an off-shell quasi-local conserved charge and it is exactly the ADT charge for any solution with any asymptotically behavior which admit the Killing vector field $\xi$. Also, as we mentioned and deduced earlier, we can calculate this charge on any time-like codimension-2 surface.
\section{Application to the BHT gravity}
\subsection{BHT gravity and its conserved charges}
The BHT gravity is a covariant gravity theory in three dimensions \cite{20}. In three dimensions, it is convenient to define dualized spin-connection and dualized curvature 2-form as
\begin{equation}\label{27}
  \omega ^{a} = \frac{1}{2} \varepsilon ^{a} _{\hspace{1.5 mm} bc} \omega ^{bc} , \hspace{1 cm} R ^{a} = \frac{1}{2} \varepsilon ^{a} _{\hspace{1.5 mm} bc} R ^{bc},
\end{equation}
respectively, where $\varepsilon _{abc}$ is Levi-Civita symbol in 3D. The Lagrangian of BHT gravity is given by (for instance, see \cite{21})
\begin{equation}\label{28}
\begin{split}
   L = & - \sigma  e^{a} \wedge R _{a} + \frac{\Lambda _{0}}{6} \varepsilon _{abc} e^{a} \wedge e^{b} \wedge e^{c} \\
     & + \frac{1}{m^{2}} \left( f^{a} \wedge R_{a} + \varepsilon_{abc} e^{a} \wedge f^{b} \wedge f^{c} \right) + h^{a} \wedge T_{a},
\end{split}
\end{equation}
where $\sigma$, $m$ and $\Lambda _{0}$ are a sign, mass parameter and the cosmological parameter, respectively. Equations of motion of the BHT gravity are
\begin{equation}\label{29}
  \begin{split}
       & E^{(e)}_{a} = - \sigma R _{a} + \frac{\Lambda _{0}}{2} \varepsilon _{abc} e^{b} \wedge e^{c} + D h_{a} - \frac{1}{2 m^{2}} \varepsilon _{abc} f^{b} \wedge f^{c} = 0 \\
       & E^{(\omega)}_{a} = - \sigma T_{a} - \frac{1}{m^{2}} D f _{a} + \varepsilon _{abc} e^{b} \wedge h^{c}=0 \\
       & E^{(f)} _{a} = - \frac{1}{m^{2}} \left( R _{a} + \varepsilon _{abc} e^{b} \wedge f^{c} \right) =0 \\
       & E^{h} _{a} = T _{a} = 0,
  \end{split}
\end{equation}
where $D$ denotes the exterior covariant derivative, and the surface term of this theory is
\begin{equation}\label{30}
  \Theta (\Phi, \delta \Phi) = - \sigma \delta \omega ^{a} \wedge e _{a} - \frac{1}{m^{2}} \delta \omega ^{a} \wedge f_{a} + \delta e^{a} \wedge h_{a}.
\end{equation}
It is obvious that this theory is torsion-free. One can solves equations of motion \eqref{29} and find that
\begin{equation}\label{31}
  h^{a}= -\frac{1}{m^{2}} C^{a}, \hspace{1 cm} f^{a}= - S^{a}.
\end{equation}
In the above equations $S_{\mu \nu} =\mathcal{R} _{\mu \nu} - \frac{1}{4} g_{\mu \nu} \mathcal{R}$ is 3D Schouten tensor and $C_{\mu \nu} = \sqrt{-g} \varepsilon _{\nu \alpha \beta} \nabla ^{\alpha} S^{\beta}_{\mu}$ is Cotton tensor, where $\mathcal{R} _{\mu \nu}$ and $\mathcal{R}$ are respectively Ricci tensor and Ricci scalar. One can calculate Eq.\eqref{15} to find $K(\xi)$ for the BHT gravity
\begin{equation}\label{32}
  K(\xi)= - \sigma (i_{\xi} \omega ^{a} - \chi _{\xi} ^{a}) e _{a} - \frac{1}{m^{2}} (i_{\xi} \omega ^{a} - \chi _{\xi} ^{a}) f _{a} + i_{\xi} e ^{a} h_{a},
\end{equation}
where $\chi _{\xi} ^{a}$ is dualized of the generator of Lorentz gauge transformations $\lambda _{\xi} ^{ab}$
\begin{equation}\label{33}
  \chi _{\xi} ^{a} = \frac{1}{2} \varepsilon ^{a} _{\hspace{1.5 mm} bc} \lambda _{\xi} ^{bc}.
\end{equation}
By substituting Eq.\eqref{30} and Eq.\eqref{33} into Eq.\eqref{18} we find that
\begin{equation}\label{34}
  \begin{split}
     Q_{ADT} (\xi) = & i_{\xi} e^{a} \delta h_{a} + i_{\xi} h^{a} \delta e_{a} - \sigma i_{\xi} e^{a} \delta \omega _{a} - \frac{1}{m^{2}} i_{\xi} f^{a} \delta \omega _{a} \\
       & - \sigma (i_{\xi} \omega ^{a} - \chi _{\xi} ^{a}) \delta e _{a} - \frac{1}{m^{2}} (i_{\xi} \omega ^{a} - \chi _{\xi} ^{a}) \delta f _{a}.
  \end{split}
\end{equation}
Since $ \lambda ^{ab} = e^{\sigma [a} \pounds _{\xi} e^{b]}_{\hspace{1.5 mm} \sigma} $ so $\chi _{\xi} ^{a}$ is given as \cite{8}
\begin{equation}\label{35}
  \chi _{\xi} ^{a} = i_{\xi} \omega ^{a} -\frac{1}{2} \varepsilon ^{a}_{\hspace{1.5 mm} bc} e^{\nu b} (i_{\xi} T^{c})_{\nu} + \frac{1}{2} \varepsilon ^{a}_{\hspace{1.5 mm} bc} e^{b}_{\hspace{1.5 mm} \mu} e^{c}_{\hspace{1.5 mm} \nu} \nabla ^{\mu} \xi ^{\nu} ,
\end{equation}
and because the BHT gravity is torsion-free then we have
\begin{equation}\label{36}
  i_{\xi} \omega ^{a} - \chi _{\xi} ^{a} = - \frac{1}{2} \varepsilon ^{a}_{\hspace{1.5 mm} bc} e^{b}_{\hspace{1.5 mm} \mu} e^{c}_{\hspace{1.5 mm} \nu} \nabla ^{\mu} \xi ^{\nu} .
\end{equation}
By substituting Eq.\eqref{36} into Eq.\eqref{19}, one can find the conserved charge of a considered solution associated to a Killing vector $\xi$.
\subsection{Rotating OTT black hole solution}
The BHT gravity admits the AdS$_{3}$ spacetime as an unique maximally symmetric background when
\begin{equation}\label{37}
  \sigma =1, \hspace{0.7 cm} m^{2}=\frac{1}{2 l^{2}}, \hspace{0.7 cm} \Lambda _{0}= -\frac{1}{2 l^{2}},
\end{equation}
where $l$ is the AdS$_{3}$ spacetime radii. Equations \eqref{36} are well-known as the BHT conditions. The rotating OTT black hole spacetime solves equations of motion of the BHT gravity when BHT conditions are satisfied. The rotating OTT black hole is defined by the metric \cite{23,24}
\begin{equation}\label{38}
  ds^{2} = - N(r)^{2} F(r)^{2} dt^{2} + F(r)^{-2} dr^{2} + r^{2} \left( d \phi + N^{\phi}(r) dt \right)^{2},
\end{equation}
where
\begin{equation}\label{39}
  \begin{split}
       & F(r)=\frac{H(r)}{r} \sqrt{\frac{H(r)^{2}}{l^2}+\frac{b}{2} H(r) (1+\eta) +\frac{b^{2} l^{2}}{16} (1- \eta)^{2} - \mu \eta }, \\
       & N(r)=1+ \frac{b l^{2}}{4 H(r)} (1-\eta), \\
       & N^{\phi}(r)=\frac{l}{2r^{2}} \sqrt{1- \eta ^{2}} \left( \mu -b H(r) \right), \\
       & H(r)=\sqrt{r^{2}- \frac{\mu l^{2}}{2} (1- \eta) - \frac{b^{2} l^{4}}{16} (1- \eta )^{2} }.
  \end{split}
\end{equation}
The metric \eqref{38} depends on three free parameters, $\mu$, $b$ and $\eta$. This metric reduces to the static OTT black hole metric when $\eta=1$ and for $b=0$, it represents the rotating BTZ black hole \cite{25}.\\
We can choose dreibein as
\begin{equation}\label{40}
  e^{0} = N(r) F(r) dt, \hspace{0.7 cm} e^{1}=F(r)^{-1} dr, \hspace{0.7 cm} e^{2}=r \left( d \phi + N^{\phi}(r) dt \right).
\end{equation}
It should be noted here that the Cotton tensor vanishes for this solution.\\
Now, we take the AdS$_{3}$ spacetime as background solution following dreibeins 
\begin{equation}\label{41}
  \bar{e} ^{0} = \frac{r}{l} dt, \hspace{0.7 cm} \bar{e} ^{1}=\frac{l}{r} dr, \hspace{0.7 cm} \bar{e} ^{2}=r  d \phi ,
\end{equation}
i.e. the AdS$_{3}$ spacetime is corresponds to $s=0$.  We take the integration surface $\Sigma $ as a circle with a radius at infinity. It can be shown that the equation \eqref{34} on $\Sigma$ reduces to
\begin{equation}\label{42}
  Q_{ADT} (\xi) = \left\{ - 2 i_{\xi} \bar{e} _{a} \delta \omega ^{a}_{\hspace{1.5 mm}\phi} + 2 l^{2} (i_{\xi} \bar{\omega} ^{a} - \bar{\chi} _{\xi} ^{a}) \delta S ^{a}_{\hspace{1.5 mm}\phi} \right\} d \phi ,
\end{equation}
for the rotating OTT black hole solution.\\
Energy corresponds to the Killing vector $\xi _{t} = \partial _{t}$. For this Killing vector on the background, Eq.\eqref{36} becomes
\begin{equation}\label{43}
  i_{\xi _{(t)}} \bar{\omega} ^{a} - \bar{\chi} _{\xi _{(t)}} ^{a}  = \frac{1}{l^{2}} e ^{a}_{\hspace{1.5 mm}\phi}.
\end{equation}
By substituting Eq.\eqref{41} and Eq.\eqref{43} into Eq.\eqref{42}, we have
\begin{equation}\label{44}
    Q_{ADT} (\xi) =  2r \{ \frac{1}{l} \delta \omega ^{0}_{\hspace{1.5 mm}\phi} + \delta S ^{2}_{\hspace{1.5 mm}\phi} \} d \phi,
\end{equation}
and by integrating over one parameter path on solution space, we find that
\begin{equation}\label{45}
\begin{split}
   \int_{0}^{1} Q_{ADT} (\xi) ds = & 2r \{ \frac{1}{l} \left[ \omega ^{0}_{\hspace{1.5 mm}\phi (s=1)} - \omega ^{0}_{\hspace{1.5 mm}\phi (s=0)} \right] \\
     & + \left[ S ^{2}_{\hspace{1.5 mm}\phi (s=1)} -S ^{2}_{\hspace{1.5 mm}\phi (s=0)} \right] \} d \phi,
\end{split}
\end{equation}
where $s=1$ is corresponds to the rotating OTT black hole solution. By expanding $\omega ^{0}_{\hspace{1.5 mm}\phi (s=1)}$ and $S ^{2}_{\hspace{1.5 mm}\phi (s=1)}$ about infinity, we have
\begin{equation}\label{46}
  \begin{split}
       & \omega ^{0}_{\hspace{1.5 mm}\phi  (s=1)} = \omega ^{0}_{\hspace{1.5 mm}\phi (s=0)}+ \frac{bl}{4} (1+\eta) - \frac{l}{16r} \left[ b^{2} l^{2} (1+\eta ^{2}) + 8 \mu \right] +\mathcal{O} (r^{-2}), \\
       & S ^{2}_{\hspace{1.5 mm}\phi (s=1)} = S ^{2}_{\hspace{1.5 mm}\phi (s=0)} - \frac{b}{4} (1+\eta) - \frac{b^{2} l^{2}}{16r} (1-\eta ^{2}) +\mathcal{O} (r^{-2}),
  \end{split}
\end{equation}
therefore the equation \eqref{45} becomes
\begin{equation}\label{47}
 \int_{0}^{1} Q_{ADT} (\partial _{t}) ds = - \left\{ \mu + \frac{1}{4} b^{2} l^{2} +\mathcal{O} (r^{-1}) \right\} d \phi.
\end{equation}
By substituting Eq.\eqref{47} into Eq.\eqref{19} and by taking $r \rightarrow \infty$, we find the energy of rotating OTT black hole
\begin{equation}\label{48}
  E = \frac{1}{4} \left(\mu + \frac{1}{4} b^{2} l^{2} \right).
\end{equation}
In this subsection, we choose the normalization factor in \eqref{19} as $c=-\frac{1}{8 \pi}$. Now, we take $\xi_{(\phi)}=\partial_{\phi}$ and it is straightforward to show that
\begin{equation}\label{49}
  i_{\xi _{(\phi)}} \bar{\omega} ^{a} - \bar{\chi} _{\xi _{(\phi)}} ^{a}  = e ^{a}_{\hspace{1.5 mm}t}.
\end{equation}
In this case, Eq.\eqref{42} reduces to
\begin{equation}\label{50}
  Q_{ADT} (\xi) = - 2r \left\{ \delta \omega ^{2}_{\hspace{1.5 mm}\phi} +l \delta S ^{0}_{\hspace{1.5 mm}\phi} \right\} d \phi ,
\end{equation}
By expanding $\omega ^{2}_{\hspace{1.5 mm}\phi (s=1)}$ and $S ^{0}_{\hspace{1.5 mm}\phi (s=1)}$ about infinity, we obtain
\begin{equation}\label{51}
  \begin{split}
       & \omega ^{2}_{\hspace{1.5 mm}\phi  (s=1)} = \omega ^{2}_{\hspace{1.5 mm}\phi  (s=0)} - \frac{bl}{4} \sqrt{1-\eta ^{2}} + \frac{l}{16r} \left[ b^{2} l^{2} (1-\eta) + 8 \mu  \right] \sqrt{1-\eta ^{2}} +\mathcal{O} (r^{-2}), \\
       & S ^{0}_{\hspace{1.5 mm}\phi  (s=1)} = S ^{0}_{\hspace{1.5 mm}\phi  (s=0)} + \frac{b}{4} \sqrt{1-\eta ^{2}} + \frac{b^{2} l^{2}}{16r} (1+\eta) \sqrt{1-\eta ^{2}} +\mathcal{O} (r^{-2}).
  \end{split}
\end{equation}
In a similar way we found energy, we can find angular momentum of the rotating OTT black hole as
\begin{equation}\label{52}
  j = l \sqrt{1-\eta ^{2}} E.
\end{equation}
Now, we want to find the entropy of the rotating OTT black hole. The BHT gravity is a Chern-Simons-like theory of gravity. In the paper \cite{8}, we have found a general formula for Chern-Simons-like theories of gravity. Using obtained formula for entropy of  black hole solutions in  generalized massive gravity theory in  \cite{8}, and notice that generalized massive gravity reduces to the BHT gravity when the mass parameter of Lorentz Chern-Simons term tends to infinity,  one can calculate entropy of black hole solutions of the BHT gravity by the following formula
\begin{equation}\label{53}
  S = \frac{1}{4} \int_{0(r=r_{h})}^{2 \pi} \frac{d \phi}{\sqrt{g_{\phi \phi}}} \left( g_{\phi \phi} - \frac{1}{m^{2}} S_{\phi \phi} \right),
\end{equation}
where $r_{h}$ is radius of Killing horizon. Killing horizon of the OTT black hole is located at $r_{h}=r_{+}$ and $r_{+}$ is given by \cite{26}
\begin{equation}\label{54}
  r_{+} = l \sqrt{\frac{1+\eta}{2}} \left( - \frac{bl}{2} \sqrt{\eta} + \sqrt{\mu + \frac{1}{4} b^{2} l^{2} } \right).
\end{equation}
By some calculations, One can show that
\begin{equation}\label{55}
  S_{\phi \phi} = - \frac{1}{2} \left( \frac{H^{2}}{l^{2}} + \frac{b}{2} H (1+ \eta) + \frac{\mu}{2} (1- \eta) - \frac{b^{2} l^{2}}{16} (1+ \eta)^{2} +\frac{b^{2} l^{2}}{4} \right).
\end{equation}
By substituting the above component of Schouten tensor into Eq.\eqref{53}, the entropy of OTT black hole can be obtained as
\begin{equation}\label{56}
  S=2 \pi l \sqrt{\frac{(1+\eta) E}{2}}.
\end{equation}
Fortunately our obtained results \eqref{48}, \eqref{52} and \eqref{56} for energy, angular momentum and entropy respectively are exactly matched with results of the paper \cite{26}. So these calculations indicate that our method for calculations of conserved charges for these types of black hole solutions works as well.

\section{Conclusion}
There are several approach to obtain mass and angular momentum of black holes for higher
curvature theories \cite{2}-\cite{12'}. The authors of \cite{6} have obtained the quasi-local conserved charges for  black holes in any diffeomorphically invariant theory of gravity. By considering an appropriate variation of the metric, they have established a one-to-one correspondence between the ADT approach and the linear Noether expressions.\\
In this paper we have found quasi-local conserved charges of covariant theories of gravity in the first order formalism.
In the first order formalism of gravity, one can use the concept of the combined Lorentz-diffeomorphism symmetry to obtain the conserved charges of a gravity theory. Using this, we have found a general formula \eqref{20} for the quasi-local conserved charges in first order formalism. In this way, we can calculate the conserved charges of any solution of a gravity theory which is not asymptotically AdS or flat spacetime. Then, we have simplified the resulting formula for the Lovelock theory of gravity. Using the provided formalism we found energy and angular momentum of the rotating OTT black hole solution of BHT gravity. Then we calculated entropy of the rotating OTT black hole using the general formula for the Chern-Simons-like theories of gravity \cite{8}. Our results for energy, angular momentum and entropy are exactly coincide on the results for these quantities which have presented in the paper \cite{26}. 
 
\section{Acknowledgments}
M. R. Setare thank Y. Obukhov and T. Jacobson for helpful comments and discussions.


\begin{thebibliography}{9}

\bibitem{2} L. F. Abbott and S. Deser, Nucl. Phys. B 195, 76 (1982); ; L. F. Abbott and S. Deser, Phys. Lett. B 116, 259 (1982).
\bibitem{1}R. L. Arnowitt, S. Deser and C. W. Misner, Gen. Rel. Grav. 40, 1997, (2008).
\bibitem{3} S. Deser and B. Tekin, Phys. Rev. D 67, 084009 (2003);
 S. Deser and B. Tekin, Phys. Rev. Lett. 89, 101101 (2002);
 C. Senturk, T. C. Sisman and B. Tekin, Phys. Rev. D 86, 124030 (2012).
\bibitem{121'}J. D. Brown and J. W. York, Phys. Rev. D47, 1407, (1993).
\bibitem{6}W. Kim, S. Kulkarni, S. H. Yi, Phys. Rev. Lett. 111, 081101, (2013).
\bibitem{2'}W. Kim, S. Kulkarni, S. H. Yi,  Phys. Rev. D 88, 124004, (2013).
\bibitem{15'}A. Bouchareb, G. Clement, Class. Quant. Grav. 24, 5581, (2007).
\bibitem{16'}S. Hyun, J. Jeong, S. A. Park, S. H. Yi, Phys. Rev. D 90, 104016 (2014).
\bibitem{17'}G. Clement, Phys. Rev. D 49, 5131, (1994).
\bibitem{18'}G. Clement, Class. Quant. Grav. 11, L115, (1994).
\bibitem{8'}O. Miskovic and R. Olea, JHEP 0912, 046, (2009).
\bibitem{9'}G. Giribet and M. Leston, JHEP 1009, 070, (2010).
\bibitem{10'}O. Hohm and E. Tonni, JHEP 1004,  093, (2010).
\bibitem{11'}S. Nam, J. D. Park and S. H. Yi, JHEP 1007, 058, (2010).
\bibitem{12'}S. Nam, J. D. Park and S. H. Yi, Phys. Rev. D 82, 124049, (2010).
\bibitem{4} M. R. Setare, H. Adami, Phys. Lett. B 774, 280 (2015).
\bibitem{5} R. M. Wald, A. Zoupas, Phys. Rev. D 61, 084027 (2000) ; R. M. Wald, Phys.Rev. D 48, 3427 (1993).
\bibitem{1'} Y. Tachikawa, Class. Quant. Grav. 24, 737, (2007).
\bibitem{7} T. Jacobson, A. Mohd,  arXiv:1507.01054 [gr-qc].
\bibitem{8} M. R. Setare, H. Adami, Nucl. Phys. B 902, 115 (2016).
\bibitem{a}T. W. B. Kibble, J. Math. Phys. 2 (1961), 212;
D. W. Sciama, On the analogy between charge and spin in general relativity, in: Recent
Developments in General Relativity, Festschrift for Infeld (Pergamon Press, Oxford;
PWN, Warsaw, 1962) pp. 415-439.
\bibitem{13} M. Blagojevic, Gravitation and Gauge Symmetries (IOP Publishing, 2002).
\bibitem{b}Y. N. Obukhov,  Int. J. Geom. Meth. Mod.
Phys. 3 (2006) 95; Y. N. Obukhov, G. F. Rubilar, Phys. Rev. D74, 064002, (2006); Y. N. Obukhov, G. F. Rubilar, Phys. Rev. D76, 124030, (2007).
\bibitem{c}M. Blagojevi´c and F. W. Hehl (eds.), Gauge Theories of Gravitation, A Reader with
Commentaries (Imperial College Press, London, 2013).
\bibitem{d}M. Blagojevic', B. Cvetkovic', arXiv:1310.8309 [gr-qc]; M. Blagojevic', B. Cvetkovic', Phys. Rev. D 92, 024047 (2015).
\bibitem{9} R. M. Wald,  J. Math. Phys. 31, 2378 (1990).
\bibitem{14} P. Baecler, R. Hecht, F. W. Hehl and T. Shirafuji, Prog. Theor. Phys. 78, 16 (1987).
\bibitem{15} T. Kawai, Prog. Theor. Phys. 79, 920 (1988).
\bibitem{16} C. M. Chen, J. M. Nester and R. S. Tung, Int. J. Mod. Phys. D 24, 1530026 (2015).
\bibitem{4'} D. Lovelock, J. Math. Phys. 12, 498 (1971).
\bibitem{10} F. Correa and M. Hassaine, JHEP 1402, 014 (2014).
\bibitem{20} E. A. Bergshoeff, O. Hohm and P. K. Townsend, Phys. Rev. Lett. 102, 201301 (2009).
\bibitem{21} M. Blagojevic, B. Cvetkovic, JHEP 1101, 082 (2011).
\bibitem{23} G. Giribet, J. Oliva, D. Tempo, and R. Troncoso, Phys. Rev. D 80, 124046 (2009).
\bibitem{24} G. Giribet and M. Leston, JHEP 09, 070 (2010).
\bibitem{25} M. Banados, C. Teitelboim, J. Zanelli, Phys. Rev. Lett. 69, 1849 (1992).
\bibitem{26} M. Blagojevic, B. Cvetkovic, arXiv:1510.00069 [gr-qc].
\end{thebibliography}
\end{document}